\documentstyle[epsfig,12pt]{article}
\begin{document}
\title{Evolution of the nucleon structure in the lightest nuclei}
\author{V.V.Burov, A.V.Molochkov, G.I.Smirnov\\
{\small\em 
Joint Institute for Nuclear Research, Dubna, Russia
}}
\maketitle
\begin{abstract}
The evolution of the nucleon structure as a function of atomic mass $A$
is considered for the first time for the lightest nuclei,
$\rm D$, $^3{\rm H}$, $^3{\rm He}$ and $^4{\rm He}$, with an approach
based on the Bethe-Salpeter formalism.
We show that the pattern of the oscillation of the structure functions ratio 
$r^A(x)$ = $F_2^A$/$F_2^{\rm N(D)}$ varies with $A$ by changing the position
of the cross-over point $x_3$ in which $r^A(x)$~=~1,
unlike the pattern for nuclei with masses $A$ $>$ 4, where only
the amplitude of the oscillation changes. In particular we find that
the pattern of $F_2(x)$ modifications is controlled with the values
$(1 - x_3)$ = 0.32 (D/N), 0.16 ($^3$He/D) and 0.08 ($^4$He/D).
The obtained results follow from the relativistic consideration of the
nuclear structure and allow us to define a whole class of modifications of the
partonic distributions in the nucleon bound in a nucleus. 
The EMC effect is explained as a particular case of the considered class.
\\[.2cm]
\end{abstract}

  {\bf 1. Introduction}\\[-0.3cm]

The understanding that nucleon structure cannot be regarded as
unrelated to nuclear structure has been the main outcome of the
European Muon Collaboration (EMC)~\cite{aub83}. The  nuclear
environment modifies the nucleon partonic structure in such a way
that the ratio of the nuclear and  deuteron structure functions,
$r^A(x)$= $F_2^A$/$F_2^{\rm D}$, deviates from unity, resembling
an oscillation with respect to the line  $r^A(x)$ = 1.
  A large variety of models were suggested in an attempt to explain
the difference between $F_2^A$ and $F_2^{\rm D}$ (c.f. Review~\cite{arne}). 
All models failed to
reproduce quantitatively  $x$ dependence of the effect, many of
them experienced difficulties in describing results obtained on
different nuclear targets. Most promising were (a)~convolution
model with the parameter of separation energy~\cite{xresc},
(b)~$Q^2$ rescaling model~\cite{close1}, (c)~QCD motivated
models~\cite{close2} and (d)~quark models~(c.f. Reviews \cite{qpm}). 
It also deserves to be recalled that as has been shown in early
publications (c.f. Refs.~\cite{prl84}), the $A$ dependence of the
ratio  $r^A(x)$ could be explained with conventional nuclear
structure considerations in the range  $A >$ 4. Still,
 following resume of Ref.~\cite{arne}, we summarize
that the origin of the EMC effect remained obscure.

 It should be underlined that the major number of models considered
the EMC effect in the deep inelastic scattering (DIS) off
medium and heavy nuclei. Very few of them went
down to $A$ = 4 and $A$ = 3 in an attempt to describe the
available data by using a quark-parton
mechanism~\cite{barsh85,jaffe,pirvar,saito},
or the conventional nuclear model for $A$ = 4~\cite{morita}. 
Not a single approach aimed at the consistent evaluation of 
the {\em evolution} of the nucleon structure modifications 
in the lightest nuclei.

On one hand, it was related with specific problems of few-body
interactions and on the other hand it was simply because the role
of the binding effects which are expected to be the strongest in
the range $A \leq$ 4 was underestimated. For instance, one finds
in a recent publication (Ref.~\cite{vantt}) the statement that
``the intrinsic structure of individual nucleons is evidently not
very much affected by nuclear binding''. The statement contradicts
the conclusions of Refs.~\cite{indu,bps} in which it is found that
the consistent inclusion of the binding effects allows to describe
the data in a wide range of $x$.

The argument that the origin of the EMC effect is closely related
to saturation of the short-range binding forces in 3- or 4-nucleon
systems was first presented in \cite{sm94,sm95}.
 Previous attempts to exploit this physics for explanation of the
effect failed because of both theoretical difficulties and a lack
of understanding on how the saturation would show up in the
observables. This excuses the statement of \cite{gomez}, that the
data on $r^A(x)$ does not directly correlate with the binding energy
per nucleon. The saturation, according to~\cite{sm95}, had to
manifest itself not in the amplitude of the oscillations, but in
the pattern of the $x$ dependence of $r^A(x)$, namely in the
positions of the three cross-over points $x_i$, $i$ = 1 --- 3, in
which $r^A(x)$ = 1. Such a pattern can be clearly seen from
re-evaluated data of SLAC \cite{gomez} and NMC \cite{ama95}. There
does not  exist any data on $r^A(x)$ in the range of $A<$~4. 
Most challenging therefore is to evaluate how the effect
evolves in the range of the lightest nuclear masses, $r^D(x) \to$
$r^{A=3}(x) \to$ $r^{A=4}(x)$. When calculated in a consistent
approach, the evolution of the nucleon structure modifications can
be traced in the range $x >$ 0.3 as a certain sequence  $x_3(A=2)
<$ $x_3(A=3) <$  $x_3(A=4)$, which is much easier to check
experimentally than the deviation of $r^A(x)$ from unity. By
tagging the pattern of the modifications with the help of $x_3$,
as is shown below, one can realize how important is a correct
evaluation of $r^D(x)$, the fact, which is often ignored in many
publications.

It is natural to assume that the progress in the understanding
both of the effect and the relations between numerous models
requires development of a consistent relativistic field theory
description of nuclei. This motivated our work on the approach
which would allow us to derive an expression for the nuclear hadronic
tensor by avoiding assumptions on the mechanisms of the nuclear
binding forces. In the present Letter, we perform derivations of
the relative
 changes in the nuclear
structure function  $F_2^A(x)$, with respect to the isoscalar
nucleon one,
  $F_2^{\rm N}(x)$ = $\frac{1}{2} (F_2^{\rm p}(x)
+ F_2^{\rm n}(x))$, where $\em p$ and $\em n$ denote the free
proton and the free neutron respectively. On the other hand, the
comparison  with experimental data can be  done only in terms of
$r^A(x)$, obtained with
 the deuteron structure function $F_2^{\rm D}(x)$ as a reference.
Therefore our final results will be presented for both cases. In
the considered range of $x$ ( 0.3 $< x <$ 0.9 ) the experiments
(see Ref.~\cite{gomez}) are consistent with no $Q^2$ dependence of
$r^A(x)$. We perform numerical calculations for a fixed $Q^2$ of
10 GeV$^2$.\\

  {\bf 2. Formalism}\\[-0.3cm]

 Our approach originates from
the Bethe-Salpeter (BS) formalism~\cite{formalism} and makes
it possible to treat nuclear binding effects by using general
properties of nucleon Green functions~\cite{deut}. The method
developed in~\cite{deut} is model independent in the sense that it
does not require any assumption about the nuclear structure except
that the nuclear fragments have a small relative energy. It has
been applied  for the derivation of the relation between $F_2^{\rm
D}(x)$ and $F_2^{\rm N}(x)$.
By extending the approach for light nuclei, $A = 3$,~$4$, we have
discovered that the consistent picture of the evolution of the
partonic structure can be obtained in the same framework without
additional assumptions.

 The hadronic part of the DIS amplitude (hadronic tensor) is related
 with the forward Compton scattering amplitude $T^{A}_{\mu \nu }$
by using the unitarity condition
\begin{equation}
W^{A}_{\mu \nu }(P,q)=\frac{1}{2\pi}{{\rm Im}}\, T^{A}_{\mu \nu
}(P,q),
 \label{impart1}
\end{equation}
and $T^{A}_{\mu \nu }$ is defined as a product of electromagnetic
currents averaged over nuclear states,
\begin{equation}
T^{A}_{\mu \nu}(P,q) = i \int d^4x e^{ i q x } \langle A|  {\rm T}
\left(J_\mu \left(x \right) J_\nu \left(0\right) \right) |A
\rangle. \label{imp3}
\end{equation}

\noindent Starting from a field theory framework one can define
the matrix element in terms of solutions of the $n$-nucleon BS
equation and $n$-nucleon Green functions with the insertion of the
$\rm T$-product of electromagnetic currents,
\begin{eqnarray}
&&\langle A,P|{\rm T}(J_{\mu}\left(x)J_{\nu} (0)\right)|A,P\rangle
=\int dz_1 \dots dz_n dz^\prime_1 \dots dz^\prime_n
\bar{\chi}_{\alpha, P}^A(z_1,\dots z_n)\nonumber\\ &&\overline
\times G_{2n+2}(z_1,\dots z_n, x, z^\prime_1,\dots z^\prime_n)
{\chi}_{\alpha, P}^A(z^\prime_1,\dots z^\prime_n),
\label{xmatrres}\end{eqnarray} where $\chi^A$ is BS vertex
function introduced to describe a nucleus in terms of virtual
nucleon states. It satisfies the homogeneous BS equation,
\begin{eqnarray}
&&{\chi}_{\alpha, P}^A(x_1,\dots x_n)= \label{BSchi}\\
&&\hspace*{-1cm}\int dz_1\dots dz_n dz^\prime_1\dots dz^\prime_n
S_{(n)}(x_1\dots x_n, z_1\dots z_n) \overline G_{2n}(z_1\dots z_n,
z^\prime_1\dots z^\prime_n) {\chi}_{\alpha, P}^A(z^\prime_1,\dots
z^\prime_n). \nonumber\end{eqnarray} 
The $\overline{G}_{2n}(z_1\dots z_n, z^\prime_1\dots z^\prime_n)$ 
term denotes the irreducible truncated $n$-nucleon
Green function which is defined as follows:
\begin{equation}
\overline{G}_{2n}(z_1\dots z_n,
z^\prime_1\dots z^\prime_n)={S_{(n)}}^{-1}(z_1\dots z_n,
z^\prime_1\dots z^\prime_n)-G_{2n}^{-1}(z_1\dots z_n,
z^\prime_1\dots z^\prime_n) ,
\end{equation}
where $G_{2n}$ is an exact $n$-nucleon Green function.
The function $S_{(n)}(z_1\dots z_n, z^\prime_1\dots z^\prime_n)$ 
is the direct product of $n$ nucleon propagators. 

The BS vertex function
depends on the variables which are describing the relative
position of nucleons inside a nucleus in the four-dimensional
space. It is obvious that the main difference in kinematics of
free and bound nucleons consists in the BS vertex and Green
function dependence on
the nucleon relative time $\tau_i$ defined as
\begin{equation}
\tau_i=\frac{1}{n}\sum\limits_{j}^{n} {x_j}_0-{x_i}_0,
\end{equation}
which is otherwise fixed. This leads to important changes in the
amplitude which describes the scattering from an
off-shell-nucleon, namely, the amplitude $T_{\mu\nu}^{\rm N}$
depends now on the  component $k_0$ of the nucleon
relative momentum~$k$.
Due to this dependence the deuteron structure
function $F_2^{\rm D}$ can be related with the isoscalar nucleon
 structure function $F_2^{\rm N}$ as~\cite{deut}
\begin{eqnarray}
\!\!\!\!\!\!\! F_2^{{\rm D}}(x_{\mbox{\tiny D}})=\int
\frac{d^3k}{(2\pi)^3} \left(F_2^{{\rm N}}(x_{\mbox{\tiny N}})
\left(1-\frac{k_3}{m}\right)-\frac{M_{{\rm D}}-2E}{m} x_{{\rm
N}}\frac{d F_2^{{\rm N}}(x_{\mbox{\tiny N}})} {d x_{\mbox{\tiny
N}}}\right)\Psi^2({\bf k}), \label{f2nr}\end{eqnarray} where $E$
is the on-mass-shell nucleon energy $E^2={\bf k}^2+m^2$, $\bf
k$ is the relative three momentum of the bound nucleons, $M_{\rm
D}$ and $m$ are the masses of a deuteron and a nucleon respectively.
 The nucleon Bjorken variable is defined as 
$x_{\mbox{\tiny N}}=x_{\mbox{\tiny D}}m/(E-k_3)$, and
$\Psi^2({\bf k})$ is an analog of the three-dimensional momentum
distribution which defines motion of the on-mass-shell struck nucleon
in the field of the off-mass-shell one. The function $\Psi^2({\bf k})$ 
is directly related to the BS vertex function of
the deuteron $\Gamma^{{\rm D}}(P,{\bf k})= 
\{\Gamma^{{\rm D}}(P,k)\}_{k_0=E-M_{{\rm D}}}/{2}$~\cite{deut}:
\begin{equation}
\Psi^2({\bf k})=\frac{m^2}{4E^2 M_{{\rm D}}(M_{{\rm D}}-2E)^2}
\overline{\Gamma}^{{\rm D}}(M_{{\rm D}},{\bf k}) \sum_s u^{s}({\bf
k})\overline{u}^{s}({\bf k})\otimes \sum_s u^{s}(-{\bf
k})\overline{u}^{s}(-{\bf k}) \Gamma^{{\rm D}}(M_{{\rm D}},{\bf
k})~.
\end{equation}
The first term in Eq.~(\ref{f2nr}) arises from the nuclear Fermi
motion, while the second one results from the relative time
dependence in the bound nucleon Green functions. The competition
of these two terms defines a deviation of the ratio $F_2^{\rm
D}/F_2^{\rm N}$ from unity, which is generally considered as a
signature of the modification of the nucleon structure. 
As is shown below by extending the approach for the three- and
four-nucleon systems, it is the relationship between similar terms
which is responsible for evolution of the partonic structure in
the lightest nuclei.

\begin{figure}[h]
\begin{center}
\mbox{\epsfxsize=0.7\hsize\epsffile{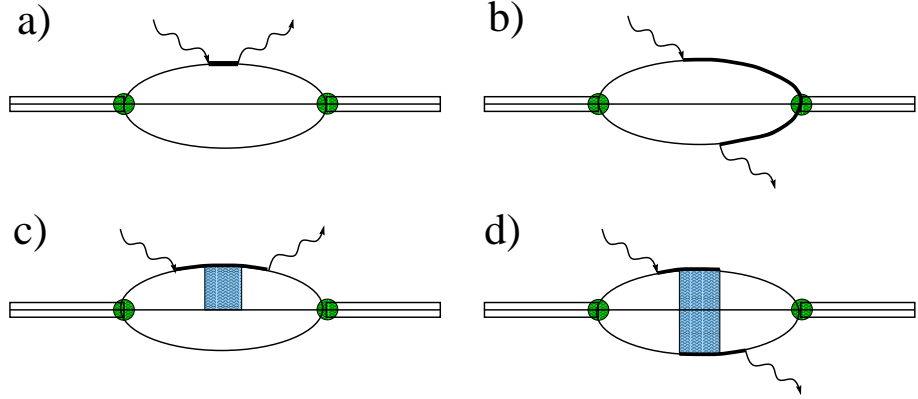}}
\end{center}

\caption{\label{diags} The diagrams which present 
 schematically the basic
contributions to the forward Compton amplitude.}
\end{figure}

Switching to momentum space we get from the Eq.~(\ref{xmatrres})
the nuclear Compton amplitude in the form:
\begin{eqnarray}
\label{compt}T_{\mu \nu}^{A}(P,q)=\int d{\cal K} d{\cal K}^\prime
\overline{\Gamma}(P,{\cal K}){S_{(n)}}(P,{\cal K})
{\overline{G}_{2(n+1)}}_{\mu\nu}(q;P,{\cal K},{\cal K}^\prime)
{S_{(n)}}(P,{\cal K}^\prime) \Gamma(P,{\cal K}^\prime),
\label{Compton}\end{eqnarray} where ${\cal K}$ denotes a set of
momenta which  describes relative motion of nucleons, ${\cal
K}=k_1,\dots, k_{n-1}$, $d{\cal K}={d^4k_1}/(2\pi)^4\dots
{d^4k_{n-1}}/(2\pi)^4$, and $P$ is the total momentum of the
nucleus.
The function $\Gamma(P,{\cal K})$ is the BS vertex function in
momentum space:
\begin{eqnarray}
S_{(n)}(P,{\cal K})\Gamma^A_\alpha(P,{\cal K})=\int d^4x_1\dots
d^4x_{n}e^{-i\sum\limits_{j=1}^{n} k_j x_j} {\chi}_{\alpha, P}(x_1\dots
x_{n}).
\end{eqnarray}
Here the Green
function ${\overline{G}_{2(n+1)}}_{\mu\nu}$ represents Compton
scattering of a virtual photon on a system of $n$-virtual
nucleons.
Typical contributions to the nuclear Compton amplitude can be
 schematically  presented by the graphs shown in Fig.~\ref{diags}.
Transpositions of virtual nucleon lines are
implied for all diagrams. Here the heavy and light lines denote 
the nucleon propagators with high and low momenta, respectively.
The graph a) represents the relativistic impulse approximation in
which only scattering off single nuclear constituents are taken
into account. The diagram b) represents contribution of
interference terms in the impulse approximation. The terms contain
the  BS vertex functions with high momenta and are suppressed as
$(1/Q^2)^l$, $l \geq 2$. The diagrams c) and d) represent
contribution of interaction corrections to
${\overline{G}_{2(n+1)}}_{\mu\nu}$. These terms contain the
contributions of two or more nucleon propagators with high momenta
and, therefore, are suppressed as $(1/Q^2)^l$.

Thus the only $Q^2$ independent term comes from the relativistic
 impulse approximation,
while irreducible interaction corrections to the imaginary part of
$T_{\mu \nu}^{A}$ are suppressed by powers of $1/Q^2$~\cite{deut}.
This justifies consideration of the
 zeroth order term of
${\overline{G}_{2(n+1)}}_{\mu\nu}$:
\begin{equation}
{\overline{G}_{2(n+1)}}_{\mu\nu}(q;P,{\cal K})=
\sum\limits_{i}{\overline G_4}_{\mu\nu}(q;P,k_i) \otimes
S^{-1}_{2n-1} (k_1,\dots k_{i-1},k_{i+1}, \dots
k_{n-1})\delta({\cal K}-{\cal K}^\prime)
 + O(1/Q^2).
\end{equation}
Then  $T_{\mu \nu}^A$ can be rewritten in terms of the
off-mass-shell nucleon Compton amplitude 
$T_{\mu\nu}^{\rm \widetilde N}=
\overline u({\bf k}_i){\overline G_4}_{\mu\nu}(q;P,k_i)u({\bf k}_i)$:
\begin{eqnarray}
T_{\mu \nu}^A(P,q)=\int d{\cal K} \sum\limits_{i}T^{\rm
\widetilde N}_{\mu\nu}(k_i,q) \overline u({\bf k}_i)S_{(n)}(P, k_i)u({\bf
k}_i) \overline{\Gamma}(P,{\cal K})S_{(n)}(P, {\cal K})
\Gamma(P,{\cal K}) . \label{compt0}\end{eqnarray}

Integration over ${k_i}_0$  can, in principle, relate $T^A_{\mu\nu}$
with on-mass-shell nucleon Compton amplitude,
$$ T^{\rm N}_{\mu \nu}(p,q) = i \int d^4x e^{ i q x } \langle {\rm N}|  {\rm T}
\left(J_\mu \left(x \right) J_\nu \left(0\right) \right) |{\rm N}
\rangle .$$ 
This can be realized
only after the singularities in nucleon propagators and the BS
vertex functions  are taken into  account~\cite{deut}. Unlike the
deuteron case, where singularities in the BS vertex function can
be neglected, in case of $A$ = 3,~4 they are connected with
nucleon-nucleon bound states, 
which lie in the range of low relative momenta.
 One can express the singularities explicitly by introducing the
``bare'' BS vertex function ${\cal G}$,
 which is regular  with respect to the relative nucleon momenta:
\begin{eqnarray}
\Gamma(P,{\cal K})=-\int d{\cal K} g_{2n}(P,{\cal K}, {\cal
K}^\prime) {S_{(n)}}(P,{\cal K}^\prime) {\cal G}(P, {\cal
K}^\prime)\label{bare},
\end{eqnarray}
where $g_{2n}$ denotes the regular part of $n$-nucleon Green
function at $P^2 \rightarrow$ $M_A^2$. This function, however,
contains singularities of $m$-nucleon Green functions ($m<n$).
 For example, in case of $^3{\rm He}$ the function  $g_6$ depends on
the exact two-nucleon propagator $G_4$, which contains a deuteron pole and
nucleon-nucleon continuous spectrum $g_4$:
\begin{equation}
G_4\left(\frac{2P}{3}+k,k_1,k^\prime_1\right)=
\frac{\Gamma^D(2P/3+k,k_1)
\overline{\Gamma}^D(2P/3+k,k^\prime_1)}
{\left(2P/3+k\right)^2 -M_D^2}
+g_4\left(\frac{2P}{3}+k,k_1,k^\prime_1\right)~.
\label{pole}\end{equation} For ${\rm ^4He}$ one has, additionally,
the ${\rm ^3He}$ and ${\rm ^3H}$ poles. Substituting 
expression~(\ref{bare}) into Eq.~(\ref{compt0}), integrating over the
relative energy of different nuclear fragments and using the 
relation~(\ref{impart1}) 
we derive the ${\rm ^3He}$, ${\rm ^3H}$ and ${\rm ^4He}$ 
hadronic tensors,
respectively, in terms of physical amplitudes of the fragments
and its derivatives over $k_0$ at the mass-shell. 

The scalar structure functions can be extracted from the hadronic
 tensors with the help of projection operators:
$$W_j^{\rm \widetilde N}(q,k_i)=
P_{j}^{\mu\nu} W_{\mu\nu}^{\rm \widetilde N}(k_i\cdot q,q^2,k_i^2),$$
where $\rm \widetilde N$ denotes the bound nucleon.
In the Bjorken limit the projection operator $g_{\mu\nu}$ can 
be used for extraction $F_2$:
$$\lim _{Q^2\rightarrow \infty }g^{\mu \nu }W^{\rm N(A)}_{\mu \nu
}(P,q) =-\frac 1x F^{\rm N(A)}_2(x)~.$$
In this case, the projection operator does not depend on the relative
momenta and the derivative of the hadronic tensor is written
as follows:
\begin{eqnarray}
&&W^{\rm \widetilde N}(k_i\cdot q,q^2,k_i^2)=
g^{\mu\nu} W_{\mu\nu}^{\rm \widetilde N}(k_i,q)\nonumber \\
&&g^{\mu\nu}\frac {d}{d{k_i}_0}
W_{\mu\nu}^{\rm \widetilde N}(k_i,q)=
\frac {d}{d(k_i\cdot q)}W^{\rm \widetilde N}(k_i\cdot q,q^2,k_i^2)
\frac{d(k_i\cdot q)}{d{k_i}_0} + 2{k_i}_0\frac
{d}{dk_i^2}W^{\rm \widetilde N}(k_i\cdot q,q^2,k_i^2).
\nonumber\end{eqnarray}
The contribution of the second term is suppressed
due to small mean value of ${k_i}_0$, and it can be neglected in the
nuclear hadronic tensor. This allows us to get rid off the dependence
of $W_{\mu\nu}^{\widetilde N}$ on $k^2_i$: 
\begin{equation}
\frac{d}{d k_0} \lim _{Q^2\rightarrow \infty }g^{\mu \nu }
W^{\rm \widetilde N}_{\mu \nu }(P,q)|_{k_0=k^{\mathrm{N}}_0} 
=\left[\frac{1}{x^2}F_2(x)-\frac{1}{x}\frac{d}{d x}F_2(x)\right] 
\left(\frac{d x}{dk_0}\right)_{k_0=k^{\mathrm{N}}_0}. 
\label{dhadr}\end{equation}
 
Introducing now Bjorken variables for a nucleus 
$x_{\mbox{\tiny\it A}}=Q^2/(2P_A\cdot q)$
and for a nucleon 
$x_{\mbox{\tiny N}}=Q^2/(2P_{\rm N}\cdot q)$, we find $F^A_2$ for
${\rm ^3He}$ and ${\rm ^3H}$ in the form:
\begin{eqnarray}
&&F_2^{{\rm ^3He}}(x_{{\rm ^3He}})=\nonumber\\
&&\hspace*{-0.7cm}
\int\frac{d^3k}{(2\pi)^3} \left[\frac{E_{\rm p}-{k_3}}{E_{\rm
p}}F_2^{\rm p}(x_{\rm p}) +\frac{E_{\rm D}-{k_3}}{E_{\rm D}}
F_2^{\rm D}(x_{\mbox{\tiny D}}) +\frac{\Delta^{{\rm ^3He}}_{\rm p}}{E_{\rm p}}
x_{\rm p}\frac{dF_2^{\rm p}(x_{\rm p})} {dx_{\rm
p}}+\frac{\Delta^{{\rm ^3He}}_{\rm p}}{E_{\rm D}}x_{\mbox{\tiny D}}
\frac{dF_2^{\rm D}(x_{\mbox{\tiny D}})}{dx_{\mbox{\tiny D}}}\right] 
{\Phi^2_{{\rm^3He}}}({\bf k}),\nonumber\\
&&F_2^{{\rm ^3H}}(x_{{\rm ^3H}})=F_2^{{\rm ^3He}}(x_{{\rm
^3He}})|_{ p\leftrightarrow n} \label{he3}
\end{eqnarray}
and for ${\rm ^4He}$ in the form:
\begin{eqnarray}
&&F_2^{{\rm ^4He}}(x_{{\rm ^4He}})=\label{he4}\\
&&\hspace*{-0.7cm}
\int\frac{d^3k}{(2\pi)^3} \left[\frac{E_{\rm p}-{k_3}}{E_{\rm
p}}F_2^{\rm p}(x_{\rm p}) +\frac{E_{\rm ^3H}-{k_3}}{E_{\rm
^3H}}F_2^{\rm ^3H}(x_{\rm ^3H}) +\frac{\Delta^{{\rm ^4He}}_{\rm
p}}{E_{\rm p}} x_{\rm p}\frac{dF_2^{\rm p}(x_{\rm p})} {dx_{\rm
p}}+\frac{\Delta^{{\rm ^4He}}_{\rm ^p}}{E_{\rm ^3H}}x_{\rm ^3H}
\frac{dF_2^{\rm ^3H}(x_{\rm ^3H})}{dx_{\rm ^3H}}\right.\nonumber\\
&&\hspace*{-0.7cm}
+\left.\frac{E_{\rm n}-{k_3}}{E_{\rm n}}F_2^{\rm n}(x_{\rm n})
+\frac{E_{\rm ^3He}-{k_3}}{E_{\rm ^3He}}F_2^{\rm ^3He}(x_{\rm
^3He}) +\frac{\Delta^{{\rm ^4He}}_{\rm n}}{E_{\rm n}} x_{\rm
n}\frac{dF_2^{\rm n}(x_{\rm n})} {dx_{\rm n}}+\frac{\Delta^{{\rm
^4He}}_{\rm ^n}}{E_{\rm ^3He}}x_{\rm ^3He} \frac{dF_2^{\rm
^3He}(x_{\rm ^3He})}{dx_{\rm ^3He}}\right] {\Phi^2_{{\rm
^4He}}}({\bf k}), \nonumber\end{eqnarray} where $\Delta^A_{\rm
N}=-M_A+E_{\rm N}+E_{A-1}$ can be interpreted
as the removal energy of the corresponding nuclear fragment. 
The three-dimensional momentum distributions $\Phi^2_{A}({\bf k})$
are defined via the ``bare'' Bethe-Salpeter  vertex functions.
For example for $^3{\rm He}$ one has:
\begin{eqnarray}
&&\Phi^2_{^3{\rm He}}({\bf k})=\frac{m M_{\rm D}}{4E_{\rm p}E_{\rm
D} M_{{\rm ^3He}}(M_{{\rm D}}-E_{\rm p}-E_{\rm D})^2} \left\{\int
\frac{d^4k_1}{(2\pi)^4}\frac{d^4k^\prime_1}{(2\pi)^4}
\overline{{\cal G}}^{^3{\rm He}}(P,k,k_1)
S_2\left(\frac{2P}{3}+k,k_1\right)\right. \label{phi}\\
&&\hspace*{-0.7cm}\times\left. \Gamma^D\left(\frac{2P}{3}+k,k_1\right)
\overline{\Gamma}^{\rm D}\left(\frac{2P}{3}+k,k^\prime_1\right)
S_2\left(\frac{2P}{3}+k,k^{\prime}_1\right) \otimes \left(\sum_s
u^{s}_\alpha({\bf k})\overline{u}^{s}_\delta({\bf k})\right)
{\cal G}^{^3{\rm He}}(P,k,k^\prime_1)\right\}_{k_0={k_0}_{{\rm
p}}}~, \nonumber\end{eqnarray} where ${k_0}_{{\rm p}}=M_{\rm
^3H}/3-E_{{\rm p}}$. Since, presently, there are no realistic
solutions of the Bethe-Salpeter equation for a bound system of
three or more nucleons, one has to use phenomenological momentum
distributions for numerical evaluations.

The momentum distribution~(\ref{phi}) describes the motion of a
nuclear constituent (N, D, ...) in the field of the off-mass-shell
spectator system. It is directly related with the nuclear momentum
distribution measured in the $e$--$A$ scattering when only
a struck nuclear constituent is detected. 
It is reasonable, thus, to assume that the momentum
distributions in Eqs.~(\ref{he3}) and (\ref{he4}) can be related
with those extracted from the experimental data. In
the calculations we make use of the distributions available
from~\cite{he3} and~\cite{new}.
The contribution arising from continuous spectra ($ppn$ for
$\rm ^3He$ and $ppnn$ for $\rm ^4He$) is small in the
considered kinematic range and does not change comparison of the
final result with the data. This justified some simplifications
which resulted in rather transparent form of Eqs.~(\ref{he3})
and~(\ref{he4}). The contributions neglected in the derivations
have been consistently taken into account in the normalization of
the momentum distributions
 $\Phi^2_{\rm ^3He}$ and $\Phi^2_{\rm ^4He}$.

This result reduces to the one obtained within the $x$-rescaling
model~\cite{xresc} and for $A$~=~3 becomes
\begin{equation}
F_2^{\rm ^3He}(x_{\rm ^3He})=\int dy d\epsilon 
\left\{F_2^{\rm p}\left(\frac{x_{\rm ^3He}} {y-\epsilon/M_{\rm ^3He}}\right) 
f^{\rm p/{^3He}}(y,\epsilon) + F_2^{\rm D}\left(\frac{x_{\rm ^3He}}
{y-\epsilon/M_{\rm ^3He}}\right)f^{\rm D/{^3He}}(y,\epsilon)\right\}~,
\label{xresc3}\end{equation} 
where $\epsilon=\Delta^{{\rm ^3He}}_{\rm p}$ has the meaning of 
a nucleon (deuteron) separation
energy and $f^{\rm p(D)/^3He}(y, \epsilon)$ are the $\rm ^3He$
spectral functions for a bound proton (deuteron):
\begin{eqnarray}
f^{\rm p(D)/^3He}(y, \epsilon)=\int \frac{d^3k}{(2\pi)^3}
\Phi^2_{^3{\rm He}}({\bf k}) \frac{m}{E_{\rm p(D)}}
y\delta\left(y-\frac{E_{\rm p(D)}-k_3}{m}\right)
\delta\left(\epsilon-(E_{\rm p}+E_{\rm
D}-M_{^3He})\right).\nonumber\\ \nonumber\end{eqnarray}

{\bf 3. Results}\\[-0.3cm]

We emphasize that both the modification of the $F_2^{\rm N}$ and its
evolution from $A$ = 1 to 4 obtained in the framework of our
method result from the relativistic consideration of the
nuclear structure.
 In the derivations we essentially exploited the fact that the
nucleons behave in a nucleus as asynchronic objects. This particular
feature is responsible for the  binding effects in $F_2^A(x)$
which arise from the dependence 
of the bound nucleon hadronic tensor on $\tau_i$.
The developed approach has
a twofold merit. First, we can naturally
reproduce the results of nonrelativistic models
(e.g.~\cite{xresc}) which offer the parametrization of the
relativistic binding effects.
Second, the outcome of the present study is particularly easy to
understand when compared with the results  of $x$-rescaling
model~\cite{xresc}. Indeed, from the comparison of
Eqs.~(\ref{dhadr}) and (\ref{he3}) with Eq.~(\ref{xresc3}) one
finds that the relative time dependence in the off-mass-shell nucleon 
Compton amplitude results in rescaling of the nucleon Bjorken $x$.
One can also notice that due to the relation existing between 
the nucleon mass and the four-dimensional radius
of its localization region, $r^2\sim 1/m^2$, the discussed here
dependence on $\tau_i$  has to lead to the increase of the
localization region of the nucleon. In a way this result resembles
the model considerations of the effect of the increase 
of the deconfinement radius or swelling of the nucleon~\cite{arne}.

The binding effects are expressed in (\ref{he3}) and (\ref{he4}) as
the first-order derivatives of nuclear fragment structure functions.
 Thus the
input structure functions $F_2^{\rm p(n)}(x)$ are responsible here
not only for the internal nucleon structure but for the dynamics
of the two-nucleon interactions as well. Similarly, 
 $F_2^{\rm D}$ is responsible for the structure
of the two-nucleon bound state and for the dynamics of
three-nucleon interactions. 
As follows from Eq.~(\ref{f2nr}), the derivative of $F_2^{\rm D}$
can be expressed in terms of the first- and second-order
derivatives of $F_2^{\rm N}$ with corresponding coefficients.
 Since the off-shell deformation of the bound deuteron structure
is determined by the second  derivative of $F_2^{\rm N}$, this
very term accounts for the three-nucleon dynamics. However, the
second  derivative of  $F_2^{\rm N}$ contributes to $F_2^{\rm
^3He}$ with a very small coefficient, $\Delta^{{\rm ^3He}}_{\rm
D}\Delta^{{\rm D}}_{\rm p}$, and the three-nucleon dynamics can
thus be neglected in the consideration of the binding effects in DIS.

The nucleon structure functions
 are introduced via
parametrizations based on the measurements of the proton and the
deuteron structure functions by DIS experiments. We used the most
recent parametrization of  $F_2^{\rm p}(x,Q^2)$ found
in~\cite{smc} and fixed the value of $Q^2$ to 10 GeV$^2$. The
structure function $F_2^{\rm n}(x)$ is evaluated from $F_2^{\rm
p}(x)$ and from the ratio  $F_2^{\rm n}(x)/F_2^{\rm p}(x)$
determined in~\cite{bcdms}. We have verified that the
uncertainties in $F_2^{\rm p(n)}(x)$ are suppressed in the
obtained ratio $r^A(x)$ and, thus, can be neglected in the
considered kinematic range. On the other hand we have checked that
an unrealistic input $F_2^{\rm N}(x)$ would have completely
destroyed the evolution of the modifications we find in the
lightest nuclei.

The results of the numerical calculations, which show how the free
nucleon structure function $F_2^{\rm N}(x)$  ($A$ = 1) evolves to
the deuteron ($A$ = 2) and helium  ($A$ = 3 and 4) structure
functions, are presented in Fig. 2(a). The evolution, which starts
from  $F_2^{\rm D}(x)$, is shown in Fig. 2(b). Contrary to what is
observed for nuclei with masses $A >$ 4, the pattern of the
oscillation of $r^A(x)$ changes its shape in the range of $A \leq$
4, which results in the change of the position of the cross-over
point $x_3$ along the line $r^A(x)$ =~1 towards higher values of $x$.

The modifications evaluated with respect to
$F_2^{\rm N}(x)$  (Fig. 2(a)) are not of academic interest only.
We use them to demonstrate that the distortions of the nucleon
structure in the deuteron cannot be regarded as negligible and,
therefore, the relation $F_2^A(x)/ F_2^{\rm D}(x)$ $\approx
F_2^A(x)/ F_2^{\rm N}(x)$ cannot be considered as justified.
Indeed, as follows from the results obtained for $A$ = 3, the
position of $x_3$ is displaced by 0.08 when $F_2^{\rm N}(x)$ is
replaced with $F_2^{\rm D}(x)$ (Fig. 2(b)). The displacement is
eight times larger than experimental error for $\overline {x_3}$
found in a recent data analysis of the ratios $F_2^A(x)/ F_2^{\rm
D}(x)$~\cite{sm99}. According to~\cite{sm99}, $\overline{x_3}$ = 0.84 $\pm$
0.01 independently of $A$ if $A >$~4. Such a precision allows one
to reliably discriminate the effect of modification of the {\em
deuteron} structure function from that of the {\em free nucleon}
structure.

It is remarkable that the value of $(1-x_3)$, which is found for
$F_2^{\rm D}(x)/F_2^{\rm N}(x)$ to be $\sim 0.32$, decreases for the
ratios $F_2^{A=3}(x)/F_2^{\rm D}(x)$ and $F_2^{\rm^4He}(x)/F_2^{\rm D}(x)$ 
to $\sim 0.16$ and $\sim 0.08$ respectively.

Further evolution of the modifications of $F_2^{\rm N}(x)$ beyond
$A$ = 4 is forbidden by Pauli exclusion principle.
As it follows from the pattern displaced in Fig.~2(a) and from the
relation between the cross-over points $x_3$, the
modifications of the nucleon structure resemble a saturation-like
process which is fully consistent with
the evolution of binding forces in the lightest nuclei.
This phenomenon allows us to introduce the
class of $x$-dependent modifications caused by the binding effects.
 Within the considered class there does not exist any 
mechanism which could
have resulted in  further changes in the pattern of $r^A(x)$
formed at the first stage of the evolution, $A \leq 4$.
The evolution of modifications to higher nuclear masses, where 
the EMC effect has been discovered, has to proceed independently
of $x$ and is to be regarded as the second stage~\cite{sm95}.
The two-stage concept of the evolution of the free nucleon structure 
in nuclear environment is decisive for the understanding of the
longstanding problem of the EMC effect.

As long as experimental data for $A$ = 2 and 3 are not available,
our predictions can be only confronted with the results on
$F_2^{\rm ^4He}(x) / F_2^{\rm D}(x)$ reported in
Refs.~\cite{gomez,ama95} and shown in Fig. 3. The position of the
cross-over point, obtained from our calculations as $x_3$ = 0.913,
is in reasonable agreement with the extrapolated data. It is of
course of high importance to improve accuracy of the data.

On the other hand, we note particularly good agreement between the
corresponding point for $A$  = 3, $x_3$ = 0.845, and the
average $x_3$-value for nuclei in the range $A = 9 \div 197$ found
in  Ref.~\cite{sm99}.
 Such an agreement naturally follows from the two-stage
concept of the $F_2^{\rm N}(x)$ evolution which is $x$ and $A$
dependent for $A \leq$ 4 and $A$ dependent only for higher masses.
The remarkable feature of our result is that the $x$ dependent
pattern of the EMC effect found experimentally in metals develops
already at $A$ = 3 and therefore can be regarded as a particular
case of the introduced here class of the modifications.\\[-0.3cm]

A fundamental relation follows from the obtained results. Since
binding corrections have the same form in Eqs.~(\ref{he3}) we can
write
\begin{eqnarray}
I ~= ~ \int\limits_0^1 \frac{dx}{x} \left(F_2^{{\rm
^3He}}(x)-F_2^{{\rm ^3H}}(x)\right)~ = \int\limits_0^1
\frac{dx}{x} \left( F_2^{\rm p}(x)-F_2^{{\rm n}}(x)\right).
\label{A3A1}\end{eqnarray} The result represents the Gottfried sum
$I$, which has often been studied experimentally  from the
combination of $F_2^{{\rm p}}(x)$ and $F_2^{{\rm
D}}(x)$(cf.~Ref.~\cite{nmcGSR}). Such a combination is equal to
$I$ to within a correction  proportional to $F_2^{\rm N}(x=0)$.
Indeed, as follows from Eq.~(\ref{f2nr}), 
$$ I_D=\int\limits_0^1
\frac{dx}{x} \left(2F_2^{{\rm p}}(x) -2F_2^{{\rm D}}(x)\right)
=~I~ - 2\frac{\langle M_{\rm D}-2E_{\rm N}\rangle_{\rm D}}{m}
F_2^{\rm N}(x=0)~.$$ Apparently, such tests cannot be performed
rigorously because $F_2^{\rm N}(x)$ is unknown at $x=0$. On the
other hand, an experiment, which used $^3$He and $^3$H targets,
would be able to measure the nucleon isospin asymmetry
independently of the model uncertainties in the binding
corrections.\\

{\bf 4. Conclusions}\\[-0.3cm]

The method for the model-free calculations of the evolution of the
nucleon structure in the lightest nuclei has been developed as the
extension of an approach based on the Bethe-Salpeter formalism.
The method allows one to express $F_2^A(x)$ in terms of structure
functions of nuclear fragments and three-dimensional momentum
distributions.

We find that the effects from asynchronic nucleons which
naturally follow from a relativistic treatment of the two-nucleon
binding are decisive in obtaining differences between structure
functions of bound and free nucleons. The characteristic
modification of the nucleon structure found for $A$ = 2  serves as
a priming for the modifications in the three- and  four-nucleon
systems and plays, therefore, a fundamental role in evolution of the
bound nucleon structure. The EMC effect, which was essentially the
observation that partonic structures of $A$ = 2 and $A$ = 56
nuclei were different, can be now regarded as a particular case of
the whole class of modifications of the free nucleon structure in
nuclear environment.

When translated to a nonrelativistic language, the event of
asynchronic nucleons can be associated  with the increase of 
the localization region for the bound nucleon 
and is observed as the modification of  $F_2^{\rm N}(x)$.

The developed approach does not require consideration of the
three-nucleon forces to describe correctly the data available for
the ratio $F_2^{\rm ^4He}$/$F_2^{\rm D}$. The two-nucleon
interactions can be, therefore, considered as the dominant mechanism
for the evaluation of the nuclear binding effects in the kinematic
range 0.3 $<x<$ 0.9.

The obtained pattern of the evolution
 of the nucleon structure function modifications
in the lightest nuclei, {\rm D}, ${\rm ^3H}$, ${\rm ^3He}$ and
${\rm ^4He}$, is consistent with saturation of the short range
binding forces. The evolution is totally different from that
observed previously for heavy nuclei, in which only the amplitude
of deviations of $F_2^A$/$F_2^{\rm D}$ from unity increased with
$A$. The quantitative predictions for ${\rm ^3He}$ and ${\rm
^4He}$ nuclei, which have to be verified in future experiments at
HERA or CEBAF, imply that the  EMC effect  in heavy nuclei can  be
naturally understood as distortions of the partonic distributions
in $^3$He or $^3$H
 which are modified by the nuclear density effects.\\

{\bf Acknowledgements}\\[-0.3cm]

We thank S.V.~Akulinichev, A.~Antonov, A.M.~Baldin, S.A.~Kulagin
and V.A.~Nikolaev for useful discussions. A.M. acknowledges the
warm hospitality of the Special Research Center for the Subatomic
Structure of Matter, Adelaide, Australia. This work was supported
in part by the RFBR grant N96-15-96423.

\small
\baselineskip=0.7\baselineskip

\begin{figure}[h]
\begin{center}
\mbox{\epsfxsize=0.7\hsize\epsffile{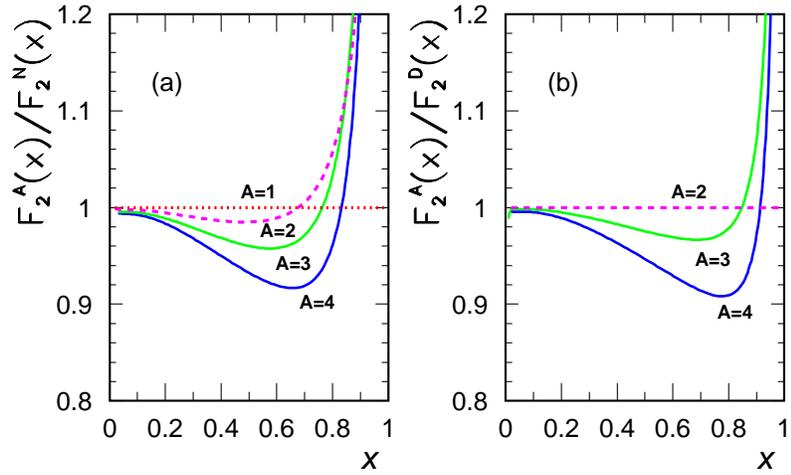}}
\end{center}

\caption{ (a) The ratio  $F_2^A(x)/F_2^{\rm N}(x)$. (b)  The ratio
of $F_2^A(x)$ to the deuteron structure function $F_2^{\rm D}(x)$
($A$ =  3, 4). $F_2^{A=3}(x)$ is defined as 
$(F_2^{\rm ^3He}(x)+F_2^{\rm ^3H}(x))/2$.
 The dashed curve in Fig. (a) shows the result of calculations,
 described in the text, for $A$ = 2. The results for $A$ = 3, 4
 are shown with the solid curves.}
\end{figure}

\begin{figure}[h]
\begin{center}
\mbox{\epsfxsize=0.7\hsize\epsffile{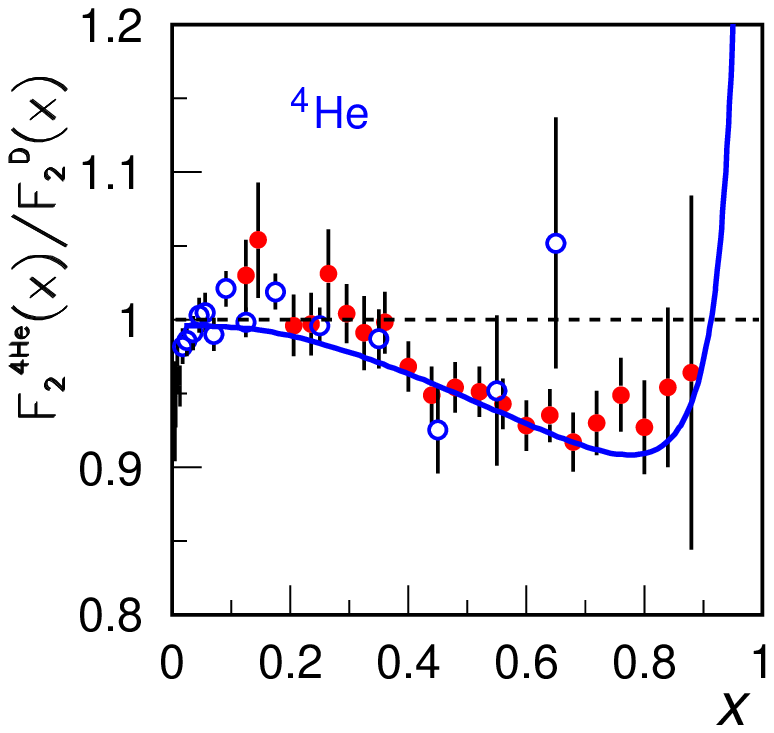}}
\end{center}

\caption{ The ratio $F_2^{\rm ^4He}(x)/F_2^{\rm D}(x)$. Results of
the calculation, described in the text, are shown with the solid
curve. The data are from Ref.~[18] (filled circles) and Ref.~[19]
(empty circles).}
\end{figure}


\begin{thebibliography}{999}
\bibitem{aub83} EMC, J.J. Aubert et al., Phys. Lett. B123, (1983), 275. 
\vspace{-0.2cm}
\bibitem{arne} M. Arneodo, Phys. Rep.  240, No. 5--6, (1994), 301.
\vspace{-0.2cm} 
\bibitem{xresc} S.V.~Akulinichev, S.A.~Kulagin and
G.M.~Vagradov, Phys. Lett. B158, (1985), 485;
S.V.~Akulinichev, Phys. Lett. B357, (1995), 451.
\vspace{-0.2cm} 
\bibitem{close1} F.E. Close, R.G. Roberts anf G.G. Ross,
Phys. Lett. B129, (1983), 346. 
\vspace{-0.2cm} 
\bibitem{close2} F.E. Close, J. Qiu and R.G. Roberts, Phys. Rev. D40, 
(1989), 2820.  
\vspace{-0.2cm}
\bibitem{qpm} C.W. Wong, Phys. Rep. 136, (1986), 1; \\
D.F. Geesman, K.Saito, A.W. Thomas,  Ann. Rev. Nucl. Part. Sci. 45, 
(1995), 337.
\vspace{-0.2cm} 
\bibitem{prl84} The earliest evaluations of the $A$ dependence in the
range $A >$ 4 and 0.2 $< x <$ 0.85 belong to R.L. Jaffe et al.,
Phys. Lett. B {\bf 134}, 449 (1984), S. Date et al., Phys. Rev.
Lett. {\bf 52}, 2344 (1984), and to C.A. Garcia Canal,
E.M. Santangelo and H. Vucetich, Phys. Rev. Lett. {\bf 53}, 1430
(1984). 
\vspace{-0.2cm} 
\bibitem{barsh85}  S. Barshay, Z. Phys. C {\bf 27} 443 (1985).
\vspace{-0.2cm} 
\bibitem{jaffe} P. Hoodboy and R.L. Jaffe, Phys. Rev. D {\bf 35}
113 (1987). 
\vspace{-0.2cm} 
\bibitem{pirvar} H. Pirner and J.P. Vary, Phys. Rev. Lett. 46 (1981) 1376.
\vspace{-0.2cm} 
\bibitem{saito} T. Uchiyama and K. Saito, Phys. Rev. C {\bf 38} 2245 (1988).
\vspace{-0.2cm}
\bibitem{morita} H. Morita and T. Suzuki, Proc. Int. School-Seminar-93,
Hadrons and Nuclei from QCD, Ed. by K. Fujii, Y. Akaishi and
B.L. Reznik, World Sci., 1993.
\vspace{-0.2cm}
\bibitem{vantt} M. V{\"a}nttinen et al., Eur. Phys. J. A {\bf 3},
351 (1998).  
\vspace{-0.2cm} 
\bibitem{indu} D. Indumathi and Wei Zhu, Z. Phys. C {\bf 74} 119 (1997).
\vspace{-0.2cm}
\bibitem{bps} O. Benhar, V.R. Pandharipande and I. Sick,
Preprint JLAB-THY-98-12, March 1998. 
\vspace{-0.2cm} 
\bibitem{sm94}  G.I. Smirnov, Phys. At. Nucl. {\bf 58}, No. 9,
1613 (1995). 
\vspace{-0.2cm} 
\bibitem{sm95}  G.I. Smirnov, Phys. Lett. B {\bf 364}, 87 (1995).
\vspace{-0.2cm}
 \bibitem{gomez} SLAC, J. Gomez et al., Phys. Rev. D {\bf 49}, 4348
(1994). 
\vspace{-0.2cm} 
\bibitem{ama95} NMC, P. Amaudruz et al., Nucl. Phys. {\bf B441}, 3
(1995). 
\vspace{-0.2cm} 
\bibitem{formalism}  E.E.~Salpeter and H.A.~Bethe, Phys. Rev. {\bf 84},
  1232 (1951).
\vspace{-0.2cm}
\bibitem{deut} V.V.~Burov and A.V.~Molochkov, Nucl. Phys. {\bf A637},
31 (1998).  
\vspace{-0.2cm}
\bibitem{he3} C.~Ciofi degli Atti and S.~Simula, Phys. Rev. C {\bf 53},
1689 (1996). 
\vspace{-0.2cm} 
\bibitem{new} R.~Schiavilla et al., Nucl. Phys. {\bf A449}, 219 (1986). 
\vspace{-0.2cm} 
\bibitem{smc} SMC, B.~Adeva et al., Phys. Lett. B {\bf 412}, 414 (1997). 
\vspace{-0.2cm} 
\bibitem{bcdms} BCDMS, A.C.~Benvenuti et al., Phys. Lett. B {\bf 237},
599 (1990).
\vspace{-0.2cm}  
\bibitem{sm99} G.I. Smirnov, hep-ph/9901422, Submitted to Eur.
 Phys. J. C.
\vspace{-0.2cm}
\bibitem{nmcGSR} NMC, M. Arneodo et al., Phys. Rev. D {\bf 50}, R1
(1994).
\end{thebibliography}
\end{document}